\documentclass[aps,prb,amsmath,amssymb,showpacs,twocolumn]{revtex4}
\usepackage{graphicx} 
\usepackage{dcolumn} 
\usepackage{array} 
\usepackage{bm} 
\usepackage{latexsym} 

\begin{document} 



\title{Diffusion of particles interacting by long-range and oscillating forces. }

\author{Filip Krzy\.zewski, Magdalena A. Za{\l}uska--Kotur}
\email{zalum@ifpan.edu.pl} 
\affiliation{Institute of Physics, Polish
Academy of Sciences, Al.~Lotnik{\'o}w 32/46, 02--668 Warsaw, Poland}

\date{\today}

\begin{abstract}

   Collective diffusion coefficient in a one dimensional lattice gas
   adsorbate is calculated using variational approach. Particles interact
   via either a long-range, or a long range electron-gas-mediated (for a
   metallic substrate), or a $12-6$ Lennard-Jones interaction. Diffusion
   coefficient as a function of the adsorbate density strongly depends on
   the relationship between the substrate lattice constant and the 
   characteristic length of the inter--particle interaction potential (which
   determines positions of the potential energy minima). The diffusion
   coefficient at fixed density as a function of the interaction 
   characteristic length has an oscillating character due to the interplay
   between the inter--particle distances allowed by the substrate lattice
   structure and the average inter--particle distances which minimize the 
   total interaction energy.
\end{abstract}



\maketitle

\section{\label{sec:A}Introduction}

Manipulation of single atoms and self-assembly techniques are very
useful methods in quantum engineering. Self-assembly is determined by
inter--atomic interactions, which at the crystal surfaces can be direct
or indirect (i.e.\ induced by the substrate). Recent studies on adatom
arrangement and their dynamics on metallic surfaces show that they
experience an indirect electron-gas-mediated interactions. Electronic
surface states are a source of a long range interactions, which decay
with the inter--particle distance $r$ as $1/r^2$ and often have an
oscillatory character
\cite{Wahlstrom,Zhang,Hyldgaard,Repp,Knorr,Tiwary,Yokoyama,Schiffrin,
  Ding}. Such interactions between adatoms lead to their
self-alignment in rows
\cite{Schiffrin,Ding,Bussmann,Negulyaev,Lagoute,Liu,Fernandez-Torrente}
or in hexagonal structures \cite{Repp,Silly,Negulyaev2}. This ordering
mechanism is a good candidate to be used for constructing and
manipulating nanostructure systems.  Linear arrangements have unique
magnetic and/or electronic properties.  The ordering dynamics and a
stability of an ordered structure depend on diffusion of adatoms on
the surface while diffusion in a system of adatoms is controlled by
interactions between them.

The collective or chemical diffusion coefficient of adsorbed species
characterizes a relaxation of the local density fluctuations in a many
particle system. It involves jumps of individual atoms (adsorbed
particles) from one binding site to another. Theoretical description
of the collective diffusion process is a complicated many--body
problem and various approaches have been applied to it, ranging from
analytic ones based on master, Fokker--Planck, or Kramers equations to
numerical Monte Carlo or molecular dynamics simulations. An important
background is provided in the works of Reed and Ehrlich\cite{reed81},
an early summary by Gomer\cite{gomer90}, and in reviews by Danani {\it
  et al.}\cite{danani97} and Ala-Nissila {\it et
  al.}\cite{alanissila02}.  The variational approach to the collective
diffusion problem, used here, was developed in a series of
works\cite{gortel04,zaluska05,badowski05,zaluska05a,zaluska06,Yakes07,
  zaluska07,krzyzewski08} and was shown to be a very efficient tool to
analyze collective diffusion problems for various types of
inter--particle interactions for homogeneous or inhomogeneous
substrates either in one or two dimensions.

We have shown in Ref. \onlinecite{zaluska06} that the long range
repulsive interactions can be responsible for rapid macroscopic
rearrangements of adatoms upon minuscule adsorbate density changes. We
have shown that the adsorbate density dependent diffusion coefficient
has peaks at densities corresponding to any ordered phase in a devil
staircase phase diagram. It is well known\cite{slavin96,devil}
that the devil staircase structure emerges when the inter--particle
interaction potential is repulsive and decays faster than $1/r$. It
has been shown, one the other hand, that adatoms on metallic surfaces
often interact via forces with oscillating in $r$ potential energies
\cite{Wahlstrom,Zhang,Hyldgaard,Repp,Knorr,Tiwary,Yokoyama,Schiffrin,
  Ding}.  They exhibit a $1/r^2$ decay modified by Friedel
oscillations of the electron gas correlation function meaning that
particles attract each other at some distances and repel at
others. Such a distance dependence of the inter--particle interactions
should, in principle, result in a diffusion character similar to that
already investigated for pure $1/r^2$ repulsion. On the other hand,
the oscillating interaction potential has local minima. Similarly, a
minimum is present also for the interaction of the Lennard--Jones
type. The question arises how the presence of such a minimum (or
minima) affects the diffusion kinetics in a many particle system. In
particular, it is an interesting question how does the oscillating
character of the interactions decaying like $1/r^2$ modify diffusion
in comparison with the interactions decaying like $1/r^2$
monotonically.

It is well known that the shape of interaction potential, in
particular, existence of its attractive parts is very important for
the static behavior of the system. Systems with attractive
interactions form stable clusters of ordered phases, whereas ordering
via repulsion is always global -- it affects the entire
system. Consequently, an ordered phase due to attractive interactions
occurs for a wider range of adatom densities so it is easier to
observe experimentally\cite{Silly}. Such difference in static
properties has to affect the dynamic behavior too, so it should affect
also the diffusion process.  For example, a fast collective diffusion
in ordered structures\cite{Yakes07,zaluska06} leads to a fast
reorganization of the adsorbed layer.

In this work we compare three types of interactions: monotonically
decaying like $1/r^2$, the oscillating ones decaying like $1/r^2$ with
electron-gas-mediated oscillations, and the $12-6$ Lennard--Jones
interactions. We analyze the influence of the shape of the interaction
potential on the diffusion coefficient at different adsorbate
densities.  Magnitude of the diffusion coefficient depends on several
parameters, most of them related to the character and strength of the
inter--particle interaction. In what follows we analyze how periodic in
$r$ variation of the potential superimposed on the $1/r^2$ decay
influences diffusion. We show that for systems of particles
interacting via oscillating electron-gas-mediated forces or via forces
corresponding to the $12-6$ Lennard--Jones potentials the diffusion
process depends sensitively on the ratio between the distance $r$ at
which the potential has a minimum and the substrate lattice
constant. This ratio allows to distinguish between the commensurate
and incommensurate type of diffusion kinetics. We show that the
diffusion coefficient depends periodically on the 
characteristic length of the potentials under investigation. 

The paper is organized as follows. In Section II the approach to the
diffusion coefficient calculation is shortly described. Section III
contains description of results for electron-gas mediated, oscillation
potential , and then for $12-6$ Lennard--Jones potential.  Section IV
summarizes main results of the work.

\section{\label{sec:B}Model}

System of $N$ particles interacting via long range forces is
distributed homogeneously over a one-dimensional substrate of length
$L$ with a lattice constant $a$. The interaction of two particles at
the lattice positions $ l_i$ and $ l_j$ contributes the potential
energy $\varepsilon(al_i-al_j)$ to the total energy of the system.
Following Ref.~\onlinecite{slavin96,zaluska06} we consider systems
with the pair potential energy $\varepsilon(r)$ decreasing rapidly
with $r$.  This justifies to neglecting the next-nearest-particle
interactions and accounting only for pair interactions between
neighbors no matter how large the intra-pair separation $a l$ is.  The
total interaction energy of the system is $\sum_l n_l \varepsilon(al)$
where $n_l$ is the number of nearest neighbors pairs of length $l$ (in
units of $a$) and only $l$'s satisfying the condition $\sum_l l n_l =
L$ are admitted in the sum.  In a grand canonical ensemble approach we
let $\ell$ to vary from $0$ to $\infty$ and keep, instead, the system
under fixed external pressure $P$ (in 1D it is just an external force)
which is determined by the condition that the mean nearest neighbor
pair length $\left< l \right>$ is equal to the inverse of the actual
coverage $\theta = N/L$.  In such case a probability of a pair of a
length $l$ is\cite{slavin96,zaluska06}
\begin{eqnarray}
  \label{eq:w10}
p_\ell (P,T) = Z_1^{-1}(P,T) e^{-\beta { \tilde \varepsilon}( l, P) },
\end{eqnarray}
where
\begin{eqnarray}
\label{eq:w11}
\tilde \varepsilon(l,P)=\varepsilon(al)+aP\l
\end{eqnarray}
and 
\begin{eqnarray}
  \label{eq:w12}
Z_1(P,T) = \sum_{l=1}^\infty e^{-\beta {\tilde \varepsilon}(l, P)}
\end{eqnarray}
is a single nearest neighbor-pair isothermal--isobaric partition function.

Eqs.~(\ref{eq:w10}) through (\ref{eq:w12}) allow to determine the
thermodynamic properties of the system. In particular, the equation of
state, relation between the coverage, pressure and temperature is obtained
by evaluating the mean nearest neighbor pair length
\begin{eqnarray}
  \label{eq:w13}
\left< l \right> = Z_1^{-1}(P,T) \sum_{l=1}^\infty l e^{-\beta
  {\tilde \varepsilon}(l, P)} = - \frac{1}{\beta a}  \left( \frac{\partial
  \ln Z_1}{\partial P}\right)_T,
\end{eqnarray}
and identifying it with $1/\theta$. In the low temperature limit the
main contributions to this sum come from one or at most two terms
only\cite{zaluska06}.

Collective diffusion of the system is modeled by a kinetic lattice gas
with the particle hopping rates depending on the actual potential
energy of the particle. The potential energy landscape is build by the
static potential due to the substrate, as experienced by a single
particle, and by interactions of the particle with its neighbors. Time
evolution of the system is controlled by a set of master rate equations
for the probabilities ${\cal P}(\{c\},t)$ that a {\it
  microstate} $\{c\}$ of a lattice gas occurs at time~$t$
\begin{eqnarray}
  \label{eq:1}
  &&\!\!\!\!\!\!\!
  \frac{d}{dt} {\cal P}(\{c\},t) \\ &&\!\!\!\!\!\!\!=
  \sum_{\{c'\}} \left[ W(\{c\},\{c'\})
  {\cal P}(\{c'\},t) - W(\{c'\},\{c\}) {\cal P} (\{c\},t) \right]. \nonumber
\end{eqnarray}
The microstate $\{c\}$ is understood as a set of variables specifying
which particular sites in the lattice are occupied and which are
not. $W(\{c'\},\{c\})$ is a transition probability per unit time
(transition rate) that the microstate $\{c\}$ changes into $\{c'\}$
due to a jump of a particle from an occupied site to an unoccupied
neighboring site. Microstates $\{c\}$ and $\{c'\}$ differ here only by
the position of a single particle, the one which jumped.  For
thermally activated jumps the hopping rate depends on the difference
between the energy of the system when the hopping particle is at an
intermediate position between the sites engaged in the jump and the
energy of the system when the particle is in its initial position. The
only contributions that do not cancel out in the difference is
the energy of the hopping particle in its initial position and its
energy in the activated state at the top of the potential barrier
which it jumps over.  For the particle hopping from the adsorption
site specified by a pair of integers $(l,s)$ (i.e.\ with the nearest
neighbors of adsorbed particles being at a distance $al$ and $as$,
respectively, to its left and right) to a neighboring site
$(l',s')= (l\pm 1, s\mp 1)$
the potential energy at the initial adsorption site is
\begin{eqnarray}
  \label{eq:w42}
  E_A = E_A^0 + \varepsilon(a l) + \varepsilon(a s),
\end{eqnarray}
where $E_A^0$ is static potential energy at given site due to the
interactions with the substrate. The hopping rate can be written as
\begin{eqnarray}
  \label{eq:w48}
  W(\{c\},\{c'\})=W^{l,s}_{l',s'} = W^0 e^{-\beta \left[ \Delta^{l,s}_{l',s'} -
    \varepsilon(a l) - \varepsilon(a s) \right]},
\end{eqnarray}
where $W^0=\nu \exp[-\beta (E_B^0-E_A^0)]$ is a hopping rate for an
isolated (i.e.\ non--interacting) particle and $\nu$ is an intrinsic attempt
frequency. $\Delta^{l,s}_{l',s'}$ is the amount by which the
potential energy $ E_B^0$ of the hopping particle at a bridge site between its
initial and the final position is modified by interactions with the
neighbors at each its side.  We parametrize a microstate $\{c\}$ as
$\{c\}=[X,\{m\}]$ by selecting one particle as a {\it reference
  particle}, denoting its lattice position as $X$ and specifying
positions $\{m\}=\{m_1,m_2,\ldots,m_{N-1}\}$ of all remaining $N-1$
particles with respect to it\cite{zaluska06}. $\{m\}$ is referred to
as a {\it configuration}. Master equations (\ref{eq:1}) are linear in
set of probabilities ${\cal P}(X,\{m\},t)$ and so their lattice
Fourier transform with respect to $X$ can be easily done. The result
is that $k$--components of the probabilities, ${\cal P}_{\{m\}}(k,t)$,
evolve in time independently of each other. The rate equations for
${\cal P}_{\{m\}}(k,t)$ can be expressed in terms of a $k$--space
microscopic rate matrix ${\mathbb M}(k)$ (with rows and columns
labelled by $\{m\}$) containing the individual hopping rates and phase
factors like $e^{\pm ika}$. Details can be found in
Refs.~\onlinecite{zaluska07,zaluska06}. The collective diffusion
coefficient is related to that eigenvalue $-\lambda_D(k) > 0$ (termed
the diffusive eigenvalue) of ${\mathbb M}(k)$ which vanishes like
$k^2$ in the limit $k \rightarrow 0$. This eigenvalue is then
estimated from above in a spirit of a variational
principle\cite{zaluska07} as
\begin{eqnarray}
  \label{eq:17}
  \lambda_D( k) \equiv \frac{ {\tilde {\bm \phi}} \cdot 
[-{\mathbb M}(k)] \cdot {\bm  \phi}}
  {{\tilde {\bm \phi}} \cdot  {\bm \phi} } \rightarrow -D  k^2,
\end{eqnarray}
where the $\rightarrow$ stands for the $k \rightarrow 0$ limit and
${\tilde {\bm \phi}}$ and ${\bm \phi}$ are variational trial
left and right, respectively, eigenvectors of ${\mathbb M}(k)$
corresponding to the diffusive eigenvalue. It has been shown
\cite{zaluska06,zaluska07} that for a homogeneous substrate the
$\{m\}$-th component of the trial left
eigenvector has the form
 \begin{eqnarray}
  \label{eq:14}
  {\tilde \phi}_{\{ m\}}( k) = 1 + \sum_{j=1}^{N-1}  e^{i k  a m_j}.
\end{eqnarray}
and that $\phi_{\{m\}} (k) = P^{eq}_{\{m\}} {\tilde \phi}_{\{ m\}}( k) $,
where $ P^{eq}_{\{m\}} $ is the probability of the configuration
$\{m\}$ in equilibrium.  

We calculate the diffusion coefficient $D$ as
a ratio (the $k \rightarrow 0$ limit is implied)
\begin{eqnarray}
  \label{eq:20}
  D=-\frac{\lambda_D}{ k^2} = \frac{{\cal M}( k)}{{\cal N}( k) k^2},
\end{eqnarray}
of the ``expectation value'' numerator 
\begin{eqnarray}
  \label{eq:21}
\nonumber 
{\cal M} ( k) &=& \sum_{\{m\},\{m'\}}^{\rm no\ rep} P^{\rm eq}_{\{m'\}}
  W_{\{m\},\{m'\}} \\ 
  &\times&
  \left| {\tilde \phi}_{\{m'\}}^*(k) - {\tilde
      \phi}_{\{m\}}^*( k) \right|^2,
\end{eqnarray}
to the ``normalization'' denominator
\begin{eqnarray}
  \label{eq:22}
  {\cal N}(k) = \sum_{\{\bar m\}} P^{\rm eq}_{\{\bar m\}} \left| {\tilde
    \phi}_{\{\bar m\}}( k) \right|^2.
\end{eqnarray}
Detailed balance condition was used to derive Eq.~(\ref{eq:21}) so
each $(\{m\},\{m'\})$ term in it accounts for transitions between
$\{m\}$ and $\{m'\}$ in either direction. Therefore, each
configuration pair $(\{m\},\{m'\})$ appears in the sum in
Eq.~(\ref{eq:21}) only once [as indicated by the comment ``no rep''
above the sum] in order to avoid double counting. In the grand
canonical ensemble approach, mentioned earlier, both ${\cal N}$ and
${\cal M}$ are functions of $P, T$, and $N$.  We note in passing that ${\cal N}$ and ${\cal M}$ are
directly related to the diffusion coefficient static (or
thermodynamic) and kinetic factor, respectively, which the diffusion
coefficient is customarily factorized into\cite{reed81,gomer90}. The
former is controlled only by the static interactions, determining the
equilibrium properties of the system, while the latter is also
sensitive to the dynamic interactions within the adsorbate and the
dynamic interactions with the substrate, both controlling the rate of
an approach to the thermodynamic equilibrium. Certain characteristic
features of the density dependence of the static factor, often being
signatures of an onset of an organization within the system, may or
may not be compensated by the features present in the kinetic factor, resulting
in the density dependent diffusion coefficient from which such
features may be absent. This issue for long range repulsive
inter--particle interaction was examined in detail in
Ref.~\onlinecite{zaluska06}.

It was shown\cite{zaluska06}that, for the one dimensional system with
long range interactions, the denominator
(\ref{eq:22}) can be expressed as \cite{zaluska06}
\begin{eqnarray}
  \label{eq:w24}
{\cal N}(k=0;P,T,N) &=& N \frac{\left<l^2\right> - \left< l
  \right>^2}{\left< l \right>^2},
\end{eqnarray}
while the numerator can be written as
\begin{eqnarray}
  \label{eq:w28}
\nonumber
&&\!\!\!\!\!\!\!\!\!\!\!\!\!\!\!{\cal M}(k;P,T,N)
\\ &=& (ka)^2 N \sum_{{l=1} \atop {s=2}}^\infty 
  W^{l,s}_{l+1,s-1} p_l(P,T) p_s(P,T).
\end{eqnarray}
The ``no rep'' restriction in Eq.~(\ref{eq:21}) results in only the rates
of jumps from the left to right to be explicitly present in
Eq.~(\ref{eq:w28}) [alternatively, expression mathematically equivalent
to (\ref{eq:w28}) with only the right--to--left jump rates explicitly
appearing in it can be used].

To evaluate ${\cal M}(k)$ using Eq.~(\ref{eq:w28}) the potential
energy correction due to interactions of the activated particle is
needed. One of the simplest models accounting for the activated
particle interactions is obtained by realizing that the particle
hopping from the adsorption site $(l,s)$ to $(l+1,s-1)$ surmounts a
potential energy barrier at a bridge site situated, approximately, at
a distance $l+\frac{1}{2}$ and $s-\frac{1}{2}$ from its nearest left
and right adsorbed particle neighbor, respectively, and by evaluating
the interaction potential energy at the bridge by using
$\varepsilon(al)$ generalized to half-integer arguments.
Consequently, the potential energy correction due to interactions of a
particle at the bridge site between $(l,s)$ and $(l+1,s-1)$ site is
\begin{eqnarray}
  \label{eq:w43}
  \Delta^{l,s}_{l+1,s-1} = \varepsilon(al+a\frac{1}{2}) +
  \varepsilon(as-a\frac{1}{2}),
\end{eqnarray}
which, used in Eq.~(\ref{eq:w48}), leads to the following hopping rate
for the left--to--right jumps
\begin{eqnarray}
  \label{eq:w45}
  W^{l,s}_{l+1,s-1} = W^0
  e^{-\beta\left[\varepsilon(a l+a \frac{1}{2}) +
      \varepsilon(a s-a \frac{1}{2}) - \varepsilon(a l) - 
    \varepsilon(a s)\right]}.
\end{eqnarray}

Using (\ref{eq:w45}) in (\ref{eq:w28}) yields the kinetic factor
\begin{eqnarray}
  \label{eq:w46}
  {\cal M}(k;P,T,N) = (ka)^2 \frac{N W^0}{[Z_1(P,T)]^2}
  \left[ \sum_{l=1}^\infty
    e^{-\beta \tilde{\varepsilon}(l+\frac{1}{2})}\right]^2, 
\end{eqnarray}
in which the definition of $\tilde \varepsilon$ in Eq.~(\ref{eq:w11})
is used. Similarly like in $Z_1(P,T)$ in Eq.~(\ref{eq:w12}) the main
contribution to the sum over $l$ in Eq.~(\ref{eq:w46}) comes from one
or at most two terms for low enough temperatures but, for a given
value of $P$, these terms may correspond to different $l$'s than those
most significant in $Z_1$. In general, the sum has to be evaluated
numerically.

\section{\label{sec:C}Results}

It has been shown that the potential energy of interactions of adatoms
adsorbed on metallic surfaces like Cu/Cu(111) \cite{Repp}, Co or Co on
Cu(111) \cite{Knorr}, Fe or Co on Ag(111) \cite{Schiffrin}, Fe/Cu(111)
\cite{Ding} and Ce/Ag(111) \cite{Silly} vary with the inter--atomic
distance $r$ as
\begin{equation}
\label{potential}
\varepsilon(r)= -\epsilon_F {\bigg (} \frac{2
  \sin(\delta_F)}{\pi}{\bigg)}^2 \frac{\sin(2q_F r +2 \delta_F)}{(q_F
  r)^2} 
\end{equation}
where $\epsilon_F$ and $q_F$ are, respectively, the Fermi energy and
the Fermi wave vector of the surface electrons. Shape of the potential
energy depends also on the Fermi--level phase shift $\delta_F$ which
for many systems is equal to $-\pi/2$. This value will be used further
in our calculations. The inter--particle distance dependence of the
interaction energy (\ref{potential}) is plotted in Fig.~\ref{fig:1}
using a dashed line and compared with a long range purely repulsive
potential energy $\varepsilon(r)=\alpha / r^2$ used in Refs.~\onlinecite
{slavin96,zaluska06}. The value of the parameter $\alpha$ used in
Fig.~\ref{fig:1} has been chosen in such a way that the repulsive
potential energy curve forms an upper envelope of
electron--gas--mediated potential energy (\ref{potential}). Strength
of both types of interactions decays with the distance like $1/r^2$,
however, whereas the forces corresponding to the $\alpha/r^2$
potential energy are repulsive at any inter atomic separation, the
potential energy (\ref{potential}) oscillates, generating attractive
forces at some inter--particle distances and the repulsive ones at
others.  Attractive forces in the system lead to the creation of
stable structures at the surface and should affect the dynamic
properties of thesystem.  We compare here the diffusion kinetics
in adsorbates with both types of inter--particle interactions and, in
addition, in systems in which the interactions correspond to the
$12-6$ Lennard--Jones potential energy
\begin{eqnarray}
\label{eq:LJ}
  \varepsilon_{LJ}(r)=4\epsilon_{LJ}\Big[\Big(\frac{\sigma}{r}\Big)^{12}-
\Big(\frac{\sigma}{r}\Big)^{6}\Big],
\end{eqnarray}
typical for interactions between neutral atoms. The parameters
$\epsilon_{LJ}$ and $\sigma$ decide about depth and position of the
single potential minimum. The Lennard--Jones potential has a repulsive
$\propto 1/r^{12}$ wall at short distances, much steeper than $1/r^2$,
as it can be seen in the Fig.~\ref{fig:1}. At larger distances this
potential is attractive and, decaying like $1/r^6$, it is weak in
comparison with any of the other two.  It will be shown, however, that
the existence of this attractive part is sufficient for the diffusion
kinetics in the system with the Lennard--Jones interactions to be
qualitatively similar at some densities to that in systems with
oscillating interactions.

\subsection{\label{sec:D}Diffusion in systems with
  electron-gas-mediated interactions}

It is known\cite{devil} that a system of particles with long-range
unscreened repulsive interactions orders at $T=0$ at densities
(coverages) given by a rational fraction (smaller than $1$) when the
interaction potential decays faster than $1/r$. The coverage plotted
against external
potential has a fractal form called a devil staircase \cite{devil}. It
is also known\cite{slavin96} that systems with interactions decaying
faster than $1/r$
but additionally screened to nearest neighbors no matter how far
they are (as described in Sec.~\ref{sec:B}) also form ordered $T=0$
phases at coverages equal to $1/n$ where $n$ is any natural
number. The ``phase diagram'' (density vs. pressure or the chemical
potential) in this case is not as complicated as that for the
unscreened interactions --- it has no fractal structure. For such a
system it is possible to calculate all static
properties\cite{slavin96} as well as to investigate fully the
collective diffusion kinetics\cite{zaluska06} for many models of
microscopic kinetics. In general, the collective diffusion coefficient
peaks at sufficiently low temperatures at densities at which the
system orders (except when sharp drops in compressibility are
compensated fully by kinetics for very special models of microscopic
kinetics\cite{zaluska06}). For these densities and for densities around
them (at finite temperatures) the diffusion is very fast, meaning that
particles have an ability to rearrange quickly when the density
changes even by a small amount.

\begin{figure}
\includegraphics[width=6.0cm, angle=270]{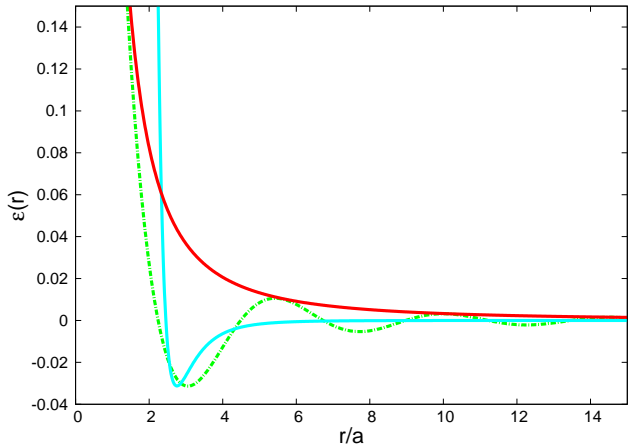}
\caption{\label{fig:1} (color online) Inter--particle interaction
  potential energies as a function of the inter--particle distance
  $r$. Thick solid line (red online): $\varepsilon(r)=\alpha/r^2$ with
  $\alpha/a^2=0.33$; dashed line (green online): the oscillating
  potential energy in Eq.~(\ref{potential}) with $q_Fa=0.7$,
  $\epsilon_F=0.36$ and $\delta_F=-\pi/2$; thin solid line (blue
  online): the $12-6$ Lennard--Jones potential energy in
  Eq.~(\ref{eq:LJ}) with $\sigma/a = 2.455$ and
  $\epsilon_{LJ}=0.031$. Energies are measured in arbitrary energy
  units. }
\end{figure}

\begin{figure}
\includegraphics[width=6.0cm, angle=270]{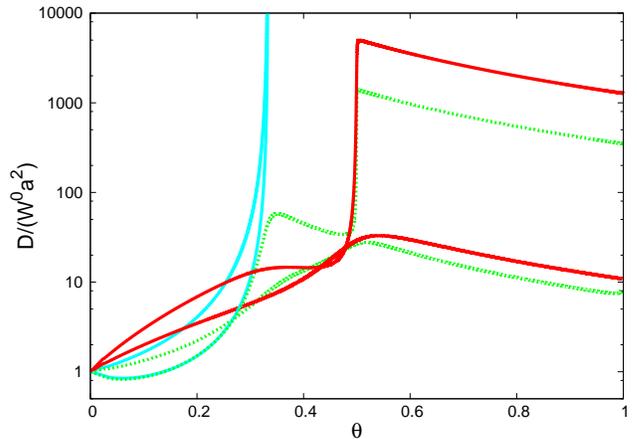}
\caption{\label{fig:2} (color online) Coverage dependence of the
  collective diffusion coefficient at temperatures $\beta=1/k_BT=60$
  (higher curves) and $\beta=20$ (lower curves) for the interaction
  potential energies plotted in Fig.~\ref{fig:1}. Thick continuous
  lines (red online): for $\varepsilon(r)=\alpha/r^2$; dashed line (green
  online): for the oscillating potential energy in
  Eq.~(\ref{potential}); thin solid line (blue online): for the $12-6$
  Lennard--Jones potential energy in Eq.~(\ref{eq:LJ}). The
  interaction potential energy parameters are the same as in
  Fig.~\ref{fig:1} and $1/\beta$ is expressed in the same arbitrary
  energy units as the interaction potential energy.}
\end{figure}

When the interactions are purely repulsive, like for
$\varepsilon(r)=\alpha/r^2$, the rearrangement into an ordered phase
occurs for the entire lattice gas at once, the phase transition is of
the second order, and formation of ordered domains surrounded by
disordered regions with lower or higher densities is not
possible. Such possibility opens for systems with interactions like in
Eq.~(\ref{potential}) which, depending on inter--particle distances
alternate between repulsion and attraction. When particles attract
each other at some distances it is possible to order the system
locally even if globally the density is too low for that\cite{
  Silly,Negulyaev}. In Fig.~\ref{fig:2} we compare the coverage
dependence of the diffusion coefficient plotted at two different
temperatures for the interaction potential energies plotted in
Fig.~\ref{fig:1} (with the same parameters as there).  It can be seen
that the character of the curves corresponding to the interaction
potential energies $\varepsilon=\alpha/r^2$ and the oscillating one
[given in Eq.~(\ref{potential})] is similar at the same
temperatures. The only significant qualitative difference between
these two cases is seen around $\theta =1/3$ at the lower of the two
temperatures considered: the diffusion coefficient for the oscillating
interaction exhibits a peak almost invisible for the system with a
purely repulsive interaction at the same temperature despite the fact
that for purely repulsive interaction the system does order at
$\theta=1/3$ at $T=0$. It indicates that the presence of the
inter--particle attraction for the system with oscillating interaction
allows for an ordering corresponding to $\theta=1/3$ already at
temperatures much higher than those are needed for such ordering with
the repulsive long range interactions only.  

In fact, $r=3a$ is the preferred inter--particle distance (based solely on
on the total interaction energy considerations) for the potential energy
parameters selected in this example, so local ordered domains are
preferentially formed which correspond to a local coverage $\theta=1/3$.
Attraction, creating the potential energy minimum, is a necessary condition
for the formation of the local ordered domains.  High diffusion coefficient
within such domains aids in their formation because collective diffusion of
particles effectively controls the system ability to create ordered
phases\cite{Repp,Knorr,Silly,Negulyaev}. With purely repulsive interactions
the ordering at $\theta=1/3$ is possible only globally so the diffusion
coefficient peak, very sharp at $T=0$, is easily smeared out by thermal
fluctuations. We can generalize our conclusions here: structures build by
attractive inter--particle interactions are capable of creating local
domains of ordered phases.  Value of the ratio of the potential energy
spatial oscillation period to the substrate lattice constant determines
which of the diffusion coefficient peaks (present for $T=0$ at $\theta=1/n$
for purely repulsive long range interactions) is amplified or attenuated.

Analyzing Fig.~\ref{fig:2} a bit more in detail for the case of lower
temperatures we see that at coverages below $\theta=0.5$ both the
oscillating and the strictly repulsive interactions result in the
diffusion coefficient of roughly the same magnitude except, however,
around $\theta=1/3$ where the broad peak for the oscillating
interactions results in a somewhat faster diffusion for this case. Both
types of interactions lead to a sharp increase of the diffusion
coefficient at $\theta=1/2$ -- another coverage of a $T=0$ ordered
phase\cite{zaluska06}. For $\theta > 0.5$ the diffusion is more
efficient for the purely repulsive interactions than it is for the
oscillating ones. This can be traced back to the fact that the purely
repulsive interaction has much stronger repulsive core at $r=a$ than
the oscillating interaction has (c.f.\ Fig.~\ref{fig:1}). All these
features are largely washed out by thermal fluctuations at the
higher temperature of the two considered here.

We show also in
Fig.~\ref{fig:2} the diffusion coefficient for the Lennard--Jones
interaction. Here, the interactions favor the inter--particle distance
of about $3a$, almost the same as the oscillating interaction does so
it is not surprising that at low concentrations both interactions
result in almost the same value of the diffusion coefficient. At
higher concentrations, however, the diffusion kinetics is controlled
by a very steep repulsive core at short distances, much steeper for
the Lennard--Jones than for any of the remaining two interactions.
 Consequently, the diffusion
coefficient increases to very high values already at
$\theta=1/3$. More features unique to the Lennard--Jones interaction
will be discussed in Sec.~\ref{sec:E}.

\begin{figure}
\includegraphics[width=7.0cm]{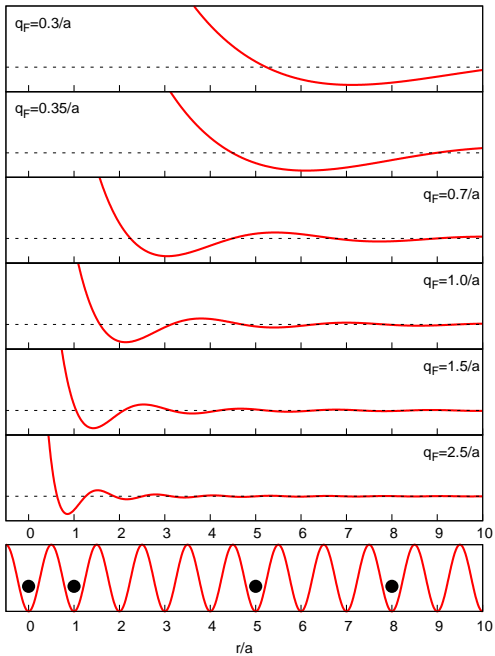} 
\caption{\label{fig:4a}(color online) Inter--particle potential energies
  (\ref{potential}) as a function of the inter--particle interaction
  distance $r$ for several values of $q_F$ (from $0.3/a$ up to
  $2.5/a$) and 
  $\epsilon_F= 0.36$ (in the same arbitrary energy units as in
  the preceding figures.)}
\end{figure}

\begin{figure}
\includegraphics[width=6.0cm, angle=270]{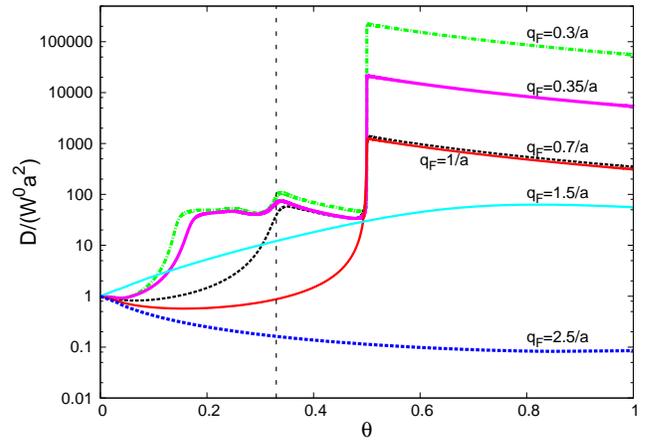} 
\caption{\label{fig:4}(color online) Density dependence of the
  diffusion coefficient in the system with oscillating inter--particle
  interaction (\ref{potential}) for several values of $q_F$ (from
  $0.3/a$ up to $2.5/a$ -- the same as in Fig.~\ref{fig:4a}),
  $\epsilon_F= 0.36$ and $\beta=60$ (the latter two parameters are
  expressed in the same arbitrary energy units as in the preceding
  figures.)}
\end{figure}

In general, the character of the density dependence of the collective
diffusion coefficient strongly depends on how closely the minima and
the maxima of the oscillating interaction potential energy match the
distances between particles which occupy the substrate lattice
sites. In Fig.~\ref{fig:4a} the interaction potential energies for
several values of $q_F$ are shown. It can be seen that the position of
the first minimum moves towards higher inter--particle distances with
decreasing $q_F$. Geometrically, the minimum inter--particle distance
possible is $a$, the substrate lattice constant, and the overall
character of the coverage dependence of the diffusion coefficient is
to a major extent determined by the character of the interaction at
this particular inter--particle distance.

If at the separation $a$ the inter--particle interaction is strongly
repulsive, as it is for $q_F=0.3/a, 0.35/a, 0.7/a$ or $1/a$, then the
diffusion coefficient behaves as a function of coverage (for coverages
between $\theta \approx 1/2$ and $1$) similarly to that corresponding
to a purely repulsive interaction (as seen already in Fig.~\ref{fig:2}
for $q_F=0.7/a$): it raises rapidly when the coverage approaches the
value $\theta=1/2$ from below, suffers a kink and then it decreases
slowly with further increase of $\theta$. This behavior was already
analyzed for a purely repulsive long--range interaction in Fig.~9 of
Ref.~\onlinecite{zaluska06}. It can be traced back to the behavior of
the diffusion coefficient static factor, proportional to ${1/\cal
  N}$ (i.e.\ proportional to an inverse of an isothermal
compressibility), modified by the behavior of the kinetic factor,
proportional to ${\cal M}$. For these values of $q_F$ the particles avoid
occupying the nearest neighbor sites so for $\theta \approx 1/2$ they
preferentially occupy every second lattice site. A substantial
additional pressure is needed to compress the system above half
occupation of the lattice, the isothermal compressibility is very low,
i.e.\ the static factor goes through a sharp
maximum at $\theta=1/2$.  The overall shape of the density dependence
of the diffusion coefficient is determined, however, by both the
static and the kinetic factor. With the hopping rates given in
Eq.~(\ref{eq:w45}) the high value for interaction potential energy at the
inter--particle distance $a$ results in a kinetic factor which also
increases sharply at $\theta=1/2$ but does not reach a maximum
there. Instead, for $\theta>1/2$ variations of the kinetic factor
almost perfectly compensate for the variations of the static factor
for these densities: the kinetic factor continues to increase (albeit,
less dramatically than at $\theta=1/2$), goes through a broad plateau
and eventually decreases sharply as $\theta$ approaches the full
coverage value of $1$. The compensation between both factors for $1/2
< \theta \le 1$ results in the diffusion coefficient which only
slowly decreases as a function of $\theta$ over this coverage interval.

The structure observed in Fig.~\ref{fig:4} at coverages below $\theta
= 1/2$ (i.e.\ at $\theta \approx 1/3$ and smaller) for $q_Fa \le 0.7$
may be understood in a similar way. In these cases the inter--particle
interaction is strongly repulsive not only at a separation $a$ but
also at the next possible one, $2a$, so at coverages close to
$\theta \approx 1/3$ the particles preferentially occupy every third
site with the resulting compressibility minimum at that coverage. We
can see in Fig.~\ref{fig:4} that locally, within the coverage interval
$0.3 < \theta < 0.47$, the behavior of the diffusion coefficient for
the interactions corresponding to $q_Fa \le 0.7$ is quite reminiscent
of that within the interval around $\theta =1/2$ and above it (for the
same values of $q_F$ and also $q_F = 1/a$): a sharp increase at
$\theta=1/3$ is followed by a slow diffusion coefficient decrease with
increasing $\theta$ (until the singularity at $\theta=1/2$ takes
over). For still lower $q_F$'s similar structures are observed in
Fig.~\ref{fig:4} around $\theta=1/6$ and even $1/7$ but due to
their overlap they are not very well resolved at the temperature
selected for the plots. 

Comparing in Fig.~\ref{fig:2} the diffusion vs.\ coverage curves
evaluated at the same temperatures for the $\varepsilon(r)=\alpha/r^2$
interaction with the ones for the corresponding to it (as defined in
Fig.~\ref{fig:1}) $q_F=0.7/a$ oscillating interaction
(\ref{potential}) we see that the presence of inter--particle
attraction for the latter results in more pronounced structures in
$D(\theta)$. The diffusion maximum around $\theta=1/3$, very broad and
almost unnoticeable for the purely repulsive interaction, becomes
quite obvious and sharp for the corresponding to it oscillating
interaction. The $\theta=1/3$ diffusion coefficient peak is clearly
visible in Fig.~\ref{fig:4} also for $q_F=0.35/a $ and $0.3/a$. The
diffusion coefficient increase at some adlayer densities (here
$\theta=1/3$) is due to the attraction felt by particles at specific
distances. The overall shape, however, of the coverage dependence of
diffusion curves is always a result of several different, often
compensating each other factors, like a character of interaction at
different distances (which affect both the static and the kinetic
factor), a relative height of the interaction potential energy barrier
which a hopping particle must overcome between adsorption sites (which
enters the kinetic factor).  It is not possible, therefore, to guess
from the interaction potential energy curves alone whether maxima of
the diffusion coefficient at certain coverages do exist or do
not. Only analyzing the diffusion coefficient vs.\ density curves we
may rationalize the existence of certain characteristic features.

Note that the structure observed around $\theta=1/3$ disappears when
$q_F$ increases from $0.7/a$ to $1/a$. Upon further increase of $q_F$
also the structure around $\theta=1/2$ disappears. Apparently, the
repulsion at the inter--particle distance of $2a$ and $a$,
respectively, ceases to be a factor controlling behavior of
diffusion. Indeed, we see in Fig.~\ref{fig:4a} that upon increasing
$q_F$ above $q_F=0.7/a$ the repulsion at an inter--particle distance of
$2a$ weakens and is gone for $q_F=1/a$ -- consequently any structure
around $\theta=1/3$ is no longer present. Then, with further increase
of $q_F$ above $q_F=1/a$ the repulsion at a separation of $a$ also
weakens and, consequently, no structure around $\theta=1/2$ is
observed for $q_F=1.5/a$.
It was checked that upon increasing $q_F$ gradually up from $0.7/a$ 
the structure around $\theta=1/3$ in Fig.~\ref{fig:4} evolves
initially into the locally concave dependence around $\theta=1/3$ and
then into the locally convex one, as seen in Fig.~\ref{fig:4} for
$q_F=1/a$. Further increase of $q_F$ washes out the only remaining
structure around $\theta=1/2$, $D(\theta)$ becomes a structureless
concave function (as seen for $q_F=1.5/a$) and then it evolves into a
convex one (as seen for $q_F=2.5/a$).
It is worth noting that the evolution of the $D(\theta)$ from a
function with the structure at $\theta=1/2$ for $q_F=1/a$ through the
concave structureless function for $q_F=1.5/a$ and then to the convex
one for $q_F=2.5/a$ parallels the evolution of $D(\theta)$ observed in
Figs.~3 and 4 of Ref.~\onlinecite{badowski05} for the short range
interaction changing from the strongly repulsive, through the weakly
repulsive, to the attractive one. Indeed, as seen in
Fig.~\ref{fig:4a}, the inter--particle interaction at a distance $a$
is strongly repulsive for $q_F=1/a$, weakly so for $q_F=1.5/a$ and
somewhat attractive for $q_F=2.5/a$.

Convex $D(\theta)$ is characteristic for systems with dominant
attractive inter--particle interactions which cause them to bond and
to form clusters. When particles stay on average at distances that
minimize the total interaction energy, the jump rates are lowered
according to (\ref{eq:w48}) resulting in a decrease of the diffusion
coefficient kinetic factor. Diffusion for $q_F=2.5/a$ oscillating
interaction decreases generally as a function of density. The
decrease, quite fast at low coverages, slows down at higher
ones. 

We have argued above that a sudden change of slope of $D(\theta)$
observed in Fig.~\ref{fig:4} at certain characteristic coverages can
be understood as due to strong inter--particle repulsion at very short
distances. This, however, cannot be the only reason for such a
behavior because, except at $\theta=1/2$, such sudden slope changes
are much less pronounced for purely repulsive interactions, certainly
the attraction must also play a role. We note also that all
$D(\theta)$ curves in Fig.~\ref{fig:4} for $q_F \le 1/a$ start as
convex functions of coverage already at $\theta=0$ and the
characteristic {\it smallest} coverage at which they suffer a sudden
decrease of its slope is $\theta_l=1/7, 1/6, 1/3$, and $1/2$ for
$q_Fa=0.3, 0.35, 0.7$, and $1$, respectively. In each case, the
corresponding value of $a/\theta_l$ is equal to the inter--particle
distance at which the potential energy (\ref{potential}) with the
appropriate value of $q_F$ has the first deepest minimum in
Fig.~\ref{fig:4a}. The convex character of $D(\theta)$ at coverages
$\theta < \theta_l$ is a signature of the attractive character of the
inter--particle interactions at distances beyond the minimum, i.e.\ the
distances larger than $a/\theta_l$. At $\theta \approx \theta_l$ the
system tends to order with inter--particle distances being, on
average, equal to $a/\theta_l$. For $\theta > \theta_l$ the average
inter--particle distances are small enough for the repulsion being the
dominant interaction and ordering is still possible at some coverages
resulting in $D(\theta)$ going through the second maximum at
$\theta=1/3$ (for $q_F \ge 0.35/a$) and $\theta=1/2$ (for $q_F \ge
0.7/a$). In short, for a particular oscillating interaction (i.e.\
particular $q_F$), the smallest coverage $\theta_l$ at which
$D(\theta)$ ceases to be a convex function is a boundary between the
coverages ($\theta < \theta_l$) for which the inter--particle
attraction dominates and the ones ($\theta> \theta_l$) for which the
repulsion does dominate. For $q_F = 2.5/a$ the interaction is
effectively equivalent to a short range attraction (the first minimum
of the interaction potential energy is at $r < a$) while for
$q_F=1.5/a$ the repulsion at $r=a$ competes with attraction at
$r=2a$. For this intermediate case the character of $D(\theta)$
resembles, as observed earlier, the one appropriate for the weak
repulsive interactions.

\begin{figure}
  \includegraphics[width=6.0cm, angle=270]{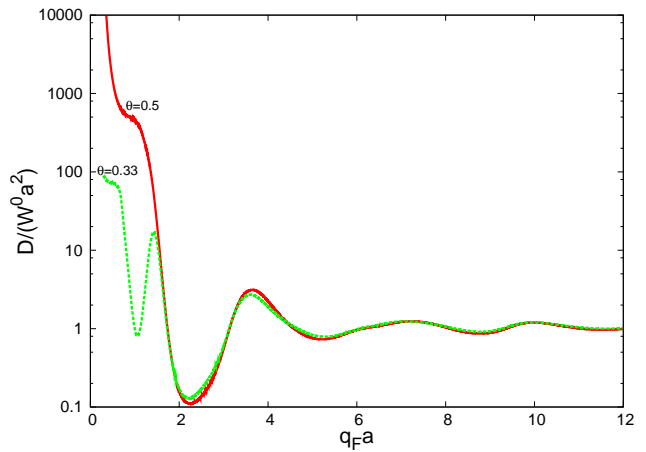}
  \caption{\label{fig:3} (color online) Diffusion coefficient $D$ for
    $\beta=60$ as a function of a parameter $q_F$ of the oscillating
    interaction potential energy (\ref{potential}) with
    $\epsilon_F=0.36$ for coverages $\theta=1/3$ (dashed line) and
    $\theta=1/2$ (solid line). $\epsilon_F$ and $\beta^{-1}$ are
    expressed in the same arbitrary energy units as in preceding
    figures.}
\end{figure}

Character of the inter--particle interaction at closest possible
separations, i.e.\ repulsion, attraction, or the potential energy
minimum, determines the overall shape of the $D(\theta)$
dependence. For small $q_Fa$'s, the particles that reside at closest
possible separations from their neighbors experience the interaction
induced repulsion. With increasing $q_Fa$ they find themselves first
at the interaction potential energy minimum, then experience
attraction and then the interaction energy maximum. The cycle repeats
with further increase of $q_Fa$. Such a cycle of successive repulsions
and attractions at a given separation should lead to a non--monotonic
dependence of the diffusion coefficient on the parameter $q_F$ at a
particular fixed coverage: we expect that after a monotonic $q_Fa$
dependence of $D$ for $q_Fa$ up to about $1$ the diffusion coefficient
should pass through a series of minima and maxima as $q_Fa$ further
increases. Indeed, this is observed in Fig.~\ref{fig:3} in which $D$
is plotted against $q_Fa$ for fixed coverage. Following the curve for
$\theta=1/3$ we observe a monotonic decrease of $D$ until a minimum is
reached for $q_Fa \approx 1$. With further increase of $q_Fa$ the
maximum is reached for $q_Fa \approx 1.4$, followed by a broad
minimum around $q_Fa \approx 2.13$, a maximum around $q_Fa \approx
3.55$, and a minimum again, at $q_Fa \approx 5$. The oscillations of
$D$ continue with increasing $q_Fa$ but their amplitude decreases
mirroring the decreasing amplitude of oscillations of the interaction
potential energy $\varepsilon(r={\rm const})$ with increasing $q_Fa$
[c.f.\ Eq.~(\ref{potential})].  For $\theta=1/2$ the $q_F$ dependence
of $D$ is very similar to that for $\theta=1/3$ except that the former starts
with much higher value for small $q_F a$, has an inflection point
around $q_Fa \approx 1.4$ rather than a narrow minimum followed by a
narrow maximum. Beyond $q_Fa \approx 2$ both curves follow closely
each other. 

\subsection{\label{sec:E}Diffusion of particles interacting via
  Lennard--Jones potential}

\begin{figure}
  \includegraphics[width=6.0cm, angle=270]{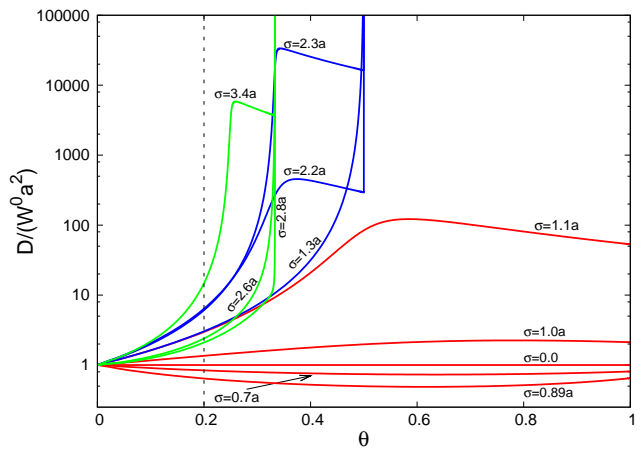}
  \caption{\label{fig:6} (color online) Coverage dependence of
    diffusion coefficient for the Lennard--Jones inter--particle
    interaction (\ref{eq:LJ}) for $\beta=20$, $\epsilon_{LJ}=0.031$
    and several values of $\sigma$. $\epsilon_{LJ}$ and
    $\beta^{-1}$ are expressed the same but arbitrary energy units.}
\end{figure}

Neutral particles are known to interact via long-range Lennard--Jones
potential energy (\ref{eq:LJ}). Its repulsive core falls off more
rapidly than $1/r^2$ (in fact, it falls as $1/r^{12}$, reaches an equlibrium
position minimum at $r_{\rm min}=2^{1/6}\sigma$, and is attractive at
larger distances, falling off like $1/r^6$).  A staircase of ordered
phases at coverages $\theta=1/n$ is supported by such an interaction
(similarly like for a purely repulsive interaction), the most
prominent ones being $\theta=1/2$ and $1/3$, and due to huge repulsion
at short distances the diffusion coefficient of particles ordered in
such phases is abnormally large in comparison to that when the system
is not ordered. 
We plot in Fig.~\ref{fig:6} the coverage dependence
of the diffusion coefficient at fixed temperature in a system with
Lennard--Jones interactions corresponding to several values of
$\sigma$ (the other parameter in Eq.~(\ref{eq:LJ}), $\epsilon_{LJ}$ is
not varied).  The overall shape of the presented curves is determined
by an interplay between two length parameters in the system: the
distance $r_{\rm min}=2^{1/6}\sigma$ below which particles repel each
other, and the substrate lattice permitted minimum separation $a$
between interacting particles.

We can systematically analyze shapes of the $D(\theta)$ curves in
Fig.~\ref{fig:6}. For $\sigma=0$ there are no inter--particle
interactions (except for site blocking) and $D(\theta)/W^0a^2= 1$ as
seen in Fig.~\ref{fig:6}. For $\sigma =0.89a$ we have $r_{\rm min}=a$ so
when $0<\sigma<0.89a$ the interaction between particles separated by
$a$ is attractive. Consequently, it is energetically preferable for
particles to be as close to each other as possible and for such values
of $\sigma$ the $D(\theta)$ dependence is convex (as seen for
$\sigma=0.7a$ and $0.89a$ in Fig.~\ref{fig:6}), typical for systems
for which the inter--particle attraction dominates (c.f.\ Fig.~4 in
Ref.~\onlinecite{badowski05}).  For $\sigma>0.89a$ the interaction at
the closest separation $r=a$ becomes repulsive and, indeed, for
$\sigma=a$, the $D(\theta)$ dependence becomes concave resembling
qualitatively $D(\theta)$ for weakly repulsive short range interaction
in Fig.~3 of Ref.~\onlinecite{badowski05}. With further increase of
$\sigma$ the repulsion at $r=a$ increases: for $\sigma=1.1a$ we note a
characteristic maximum of $D$ around $\theta=1/2$ also observed
already for stronger short range interactions in Fig.~3 of
Ref.~\onlinecite{badowski05}. For $\sigma=1.3a$ the repulsion at $r=a$
is strong enough to induce a preferential occupation of every second
site (note that the interaction at the $r=2a$ separation is still
weakly attractive in this case) causing the diffusion coefficient to
suddenly raise by many orders of magnitude as $\theta$ approaches
$1/2$ from below.

With further increase of $\sigma$ we reach, at $\sigma=2a/2^{1/6} =
1.78a$, the point beyond which the interaction at $r=2a$ is no longer
attractive. For $\sigma$ somewhat larger than $1.78a$ we have a very
strong repulsion at $r=a$ and a much weaker one at $r=2a$. We see in
Fig.~\ref{fig:6} that for $\sigma=2.2a$ a local maximum of $D(\theta)$
develops for $\theta \approx 1/3$ hinting at a preferential occupation
of every third site around this coverage due to repulsion for $r=2a$,
followed by a sharp, almost discontinuous raise of $D$ at $\theta=1/2$
due to an extremely strong repulsion at $r=a$. With further increase
of $\sigma$ the repulsion at $r=2a$ becomes stronger so the structure
around $\theta=1/3$ becomes as sharp as that around $\theta=1/2$. The
examples are curves for $\sigma=2.3a, 2.6a$ and $2.8a$ in
Fig.~\ref{fig:6}. For the latter, however, the interaction at the
separation $r=3a$ is also repulsive. In fact, with further increase of
$\sigma$ features similar to those around $\theta=1/2$ and $1/3$
develop also around $\theta=1/4$, as seen for $\sigma=3.4a$. This is
because for $\sigma> 3a/2^{1/6} = 2.67a$ the interaction at the
separation $r=3a$ becomes repulsive and, when strong enough, it leads
to a preferential occupation of every fourth site for $\theta \approx
1/4$. This structure is not present yet for $\sigma=2.8a$
at the temperature selected in Fig.~\ref{fig:6}.

\begin{figure}
\includegraphics[width=6.0cm, angle=270]{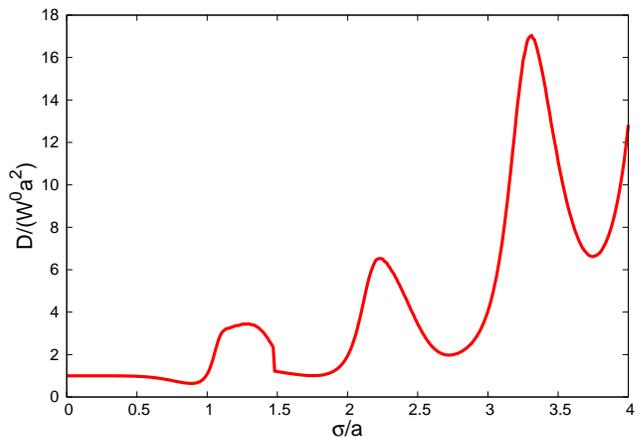}
\caption{\label{fig:5}(color online) Dependence of the diffusion
  coefficient at $\theta=1/5$ on the Lennard--Jones interaction
  parameter $\sigma$ for $\beta=20$ and
  $\epsilon_{LJ}=0.031$ (same as in Fig.~\ref{fig:6}).}
\end{figure}

The character of the coverage dependence of the diffusion coefficient
changes qualitatively every time when the interaction between
particles becomes repulsive at any of the inter--particle separations
permitted by the substrate lattice: i.e.\ every time when $\sigma$
increases through $na/2^{1/6}$. For $\sigma$ such that $n/2^{1/6} <
\sigma/a < (n+1)/2^{1/6}$ the particles separated by distances $r=a,
2a,\ldots, na$ repel each other (with the repulsion being stronger for
shorter separations) while they attract each other for separations
$r=(n+1)a$ and larger. In such case one expects at $T \approx 0$ a sharp
increase of the diffusion coefficient at coverages $\theta=1/2,
1/3,\ldots, 1/(n+1)$ due to the low temperature structural
organization of the adsorbate at these coverages to minimize the total
interaction energy in the system. In our runs done at lowest
temperatures for which the calculations are feasible we see such
structures for $\theta=1/2$, $1/3$ and $1/4$. Still, one expects that as
the parameter $\sigma$ is varied one should observe a non monotonic
oscillatory changes of the diffusion coefficient at low enough
coverages. Indeed, this is seen in Fig.~\ref{fig:5} in which
$D(\theta=1/5)$ is plotted as a function of $\sigma/a$. The maxima are
noted for $\sigma/a$ approximately {\it half way} between $\sigma=na/2^{1/6}$
for $n=1,2,3,4,\ldots$, i.e.\ for $\sigma/a \approx 1.3, 2.2, 3.1,
\ldots$, for which the interaction is already strongly
repulsive for all inter--particle separation up to $a, 2a, 3a, \ldots$,
respectively.

\section{\label{sec:F}Conclusions}

Variational approach has been applied to examine collective diffusion
in a one dimensional lattice gas systems with two type of long--range
inter--particle interaction: the electron--gas--mediated interaction
described by the oscillating Friedel--like potential energy
(\ref{potential}), and the Lennard--Jones interaction corresponding to
the potential energy (\ref{eq:LJ}). We have discussed the features of
the coverage (adsorbate density) dependence of the diffusion
coefficient for both these interactions and compared them with those
investigated in detail for purely repulsive long--range
interaction\cite{zaluska06} corresponding to the potential energy
$\propto 1/r^2$ as well as with the behavior typical for short--range
repulsive and attractive interactions\cite{badowski05}. 

In general, at densities above half coverage ($\theta>1/2$), at which
the inter--particle repulsion at short distances plays a main role,
the diffusion coefficient for the repulsive and the oscillating
interaction behaves, as a function of coverage, qualitatively
similarly: the diffusion coefficient is much higher than without
interactions and depends on coverage relatively weakly. This is true
even for the Lennard--Jones interaction, except that a very steep
repulsive core in this case, making creation of a high density
adsorbate energetically very costly, results in huge values of the
diffusion coefficient at such densities.

At lower densities, the behavior of the coverage dependence of
diffusion coefficient for the Lennard--Jones interaction, repulsive
for inter--particle separations $r<r_{\rm min}$ and attractive for
$r>r_{\rm min}$, is quite easy to understand. At sufficiently low
temperatures, it experiences for increasing $\theta$ a finite series
of progressively sharper increases at ``critical'' coverages
$\theta=1/n$ ($n=n_{\rm max}, n_{\rm max}-1, \ldots,2$). The first
one, for $\theta=1/n_{\rm max}$ corresponds to the largest
inter--particle distance $r=n_{\rm max}a < r_{\rm min}$ at which the
interaction between the particles is still repulsive (i.e.\ at the
separation $r=(n_{\rm max}+1)a > r_{\rm min}$ the interaction is
already attractive). As temperature increases, the structures at lower
densities of this series are usually smoothed out. In fact, we observe
for the Lennard--Jones interaction a delicate interplay between two
length scales: the lattice constant $a$ which determines what actual
distances between particles are possible, and the characteristic
length of the interaction potential, $r_{\rm min}$, separating the
short range repulsion from the attraction at larger distances. When
$r_{\rm min} < a$ then the repulsive core of the interaction is
irrelevant and the diffusion coefficient exhibits features similar to
those observed for short--range attractive interactions.

Features of diffusion for the oscillating interaction are somewhat
more difficult to explain in simple terms due to the existence of
multiple interaction potential energy minima and, consequently,
alternating regions of inter--particle attraction and
repulsion. Still, a sudden raise of the diffusion coefficient at
$\theta=1/2$ has origin similar to that for such structure for the
purely repulsive and Lennard--Jones interaction. The structures for
$\theta<1/2$ are, however, masked by the influence which attraction
alternating with repulsive interactions has on diffusion. 

We must note also that interpreting features of the diffusion
coefficient in terms of the features of the inter--particle
interaction is deficient in that respect that it necessarily is
limited to the interpretation of the kinetic phenomenon (diffusion) in
terms of the static properties of the system, i.e.\ in terms of the
static (or thermodynamic) factor commonly used in theories of
diffusion. The kinetic factor is known, however, to be capable of
compensating for very often drastic behavior of the static factor as a
function of density \cite{zaluska06}. This is also the case here: a slow
variation of the diffusion coefficient as a function of coverage for
several curves in Figs.~\ref{fig:4} and \ref{fig:6} for $\theta$
immediately larger than $1/2, 1/3$, and $1/4$ is the result of a
delicate compensation between a sharp drop of the static factor (which
at these particular coverages suffers a sharp cusp--like maximum
corresponding to a sharp drop in an isothermal compressibility) and a
strong increase of the kinetic factor when the coverage is increased through
these values.

\begin{acknowledgments}
  This work was supported by Poland's Ministry of Science and Higher
  Education Grant No.~N202 042 32/1171. The authors thank
  Dr.~Z.~W.~Gortel from the University of Alberta for continuing
  interest in the subject, discussions, and help in preparation of
  this manuscript.
\end{acknowledgments}

\end{document}